
\PHYSREV
\overfullrule0pt

 \newtoks\slashfraction
 \slashfraction={.13}
 \def\slash#1{\setbox0\hbox{$ #1 $}
 \setbox0\hbox to \the\slashfraction\wd0{\hss \box0}/\box0 }

  \def\Buildrel#1\under#2{\mathrel{\mathop{#2}\limits_{#1}}}

\def\lozenge{\boxit{\hbox to 1.5pt{%
             \vrule height 1pt width 0pt \hfill}}}

\def\H{{\cal H}}
\def\K{{\cal K}}
\def\L{{\cal L}}

 \doublespace
 \pubnum{6188}
\date{May   1993}
 \pubtype{T/E}
 \titlepage
\vfill
\title{Parity-Conserving  Light-Cone Quantization   of
  Quantum Field Theories
\doeack}
\author{ Ovid C. Jacob
}
\SLAC
\abstract
Parity violation is a long standing problem in light-cone quantization.
\REF\CPT{D. Soper, SLAC-REP-137, 1970, Chap. I . }
\refend
We propose a new quantization on the light-cone which treats both the
$x^{+}$ and the $x^{-}$  coordinates as light-cone 'times.'This
quantization  respects  both
parity and time-reversal. We find that now both  $P^{-}$ and $P^{+}$
become  dynamical.
\vfill
\centerline{Submitted to Physics Letters B}
 \endpage
\chapter{Introduction}
Quantizing field theories on the light-cone
\REF\dirac{ P.A.M. Dirac, Rev. Mod. Phys.\us{21},392 (1949).}
\refend
\REF\bsdisc{Ovid C. Jacob, 'Discrete Symmetries for the Bound State
Problem of Positronium in Light-Cone Quantization,' SLAC-PUB,   June
1992  .}
\refend
\REF\dba{P. A. M. Dirac, Can. Jour. Math., \us{1}, 1, (1950); 'Lectures
in Quantum Mechanics', Benjamin, NY, 1964;
P. G. Bergmann, Helv. Phys. Acta (Suppl.), \us{4}, 79 (1956).}
\refend
\REF\phitree{
T. Heinzl, S. Krusche and E. Werner, 'Spontaneous Symmetry Breaking
 in Light Cone  Quantum Field Theory' Phys. Lett. B, \us{272},
 54, (1991).}
\refend
has been done for over twenty
years. Usually, one picks the $x^{+}$ as the light-cone time, and
most of the work is in constructing the 'correct' Hamiltonian $P^{-}$
which gives the evolution in the light-cone variable $x^{+}$.  There has
been a question which always bothered us, and that is, why is $x^{+}$
'special' ? Why not pick $x^{-}$ as the light-cone time ?
Very simple considerations show that this is not such a silly question.
If we just use the simple definition of these variables
$$ x^{+} = x+t, x^{-} = x-t $$
then we find that under time reversal, ($t\rightarrow -t $),
$x^{+}\rightarrow x^{-}$ and $x^{-}\rightarrow x^{+}$ .
Parity likewise mixes these two variables : under parity
($x\rightarrow -x$), $x^{+}\rightarrow -x^{-}$ and
$x^{-}\rightarrow -x^{+} $. This suggests that we  should
treat these variables on equal footing.

There are stronger reasons though, coming from the study of the
equations of motion which the solution needs to satisfy.
The outline of this paper is as follow :
first we'll compare the equations
of motion we get in a Lagrangian formulation with those in Hamiltonian
formulation; then
we will the show how to quantize on the light-cone using
$\us{both}$ $x^{+}$ and $x^{-}$ as light-cone 'times '
(A similar situation occurs in  the quantization of $N=2$ strings
\REF\nm{Neil Marcus, Yaron Oz, Shimon Yankielowicz, 'Harmonic Space,
Self-Dual Yang-Mills and the N=2 String', TAUP-1928-91,
hep-th@p112010; Neil Marcus, 'A Tour through N=2 Strings', TAUP-2002-92,
hep-th@9211059}
\refend
\REF\ov{H. Ooguri and C. Vafa, Nucl. Phys. \us{B361}, 469(1991);
\us{B367}, 83(1991)}
\refend  ).
\chapter{Simple Considerations  on Differential Equations}
Let us study, for concreteness, $\phi^{4}$ in
1+1-dimensions. We will first study this in equal-time quantization.

The Lagrangian is
$$ \L = \partial_{t}\phi \partial_{t}\phi - \partial_{x}\phi \partial_{x}
 \phi -{\lambda \over 4!} \phi^{4}$$
This leads to the following hyperbolic
\REF\ch{R. Courant and D. Hilbert, 'Methods of Mathematical Physics,'
vol. II, Chap. III, John Willey and Sons, New York, 1962 .}
\refend
second order differential equations of motion :
$$ (\partial_{t}^{2}-\partial_{x}^{2})\phi ={\lambda \over 3!}
\phi^{3}$$
Integrating out this equation generates two constants which can be
determined by two appropriate boundary conditions.

The usual path from the classical theory to the quantum theory is to
construct the canonical pair  of field $\phi$ and its sister momentum
and to impose equal-time commutation relation on the Poisson bracket.
Note that the Lagrangian formulation gives us a second order hyperbolic
differential equation with two integration constants after integration
over time.  This integration over time gives us the time evolution of
the system.

Let us see what
happens when we go to the Hamiltonian picture.
Here, by defining the momentum $\pi=d\phi/dt$, and by defining
a Hamiltonian
we transform the above second order differential equation into set(s)
of coupled linear differential equations. In this case,
$${d\phi \over dt} = \{ \H, \pi \}$$
$${d\pi \over dt} = -\{ \H, \phi \}$$
where  $\H$ is the Hamiltonian. Note that there are still two integration
constants, one  from each linear differential equation when we integrate
over time.
This means that there are still two boundary conditions to be specified.
That's reassuring, since this means that we have the same physics as
 in the Lagrangian case since the two boundary conditions from the
second order equation  can  describe the same physics as two boundary
conditions coming from two first order differential equations.
This is all probably well-known to  the reader. We just repeat it here
because this is not always taken into account when we come to light-cone
quantization.
\chapter{Equal-Time Quantization and Parity }
Let us now define the usual
energy-momentum tensor $T^{\mu\nu}$ in the
equal-time quantization case :
$$T^{\mu\nu} = \partial ^{\mu} \phi \partial ^{\nu} \phi - g^{\mu\nu}
\L $$
and
$$P^{\mu} = \int dx^{0} T^{\mu 0}$$
In our case this means
$$\H=P^{0} = \int dx^{0} ( (\partial_{x} \phi)^{2} + {\lambda \over
4! } \phi^{4} ) $$
which is dynamic, and
$$P^{1}= \int dx^{0} (\partial_{x}\phi \partial_{t} \phi )$$
which is kinematic i.e., free of interactions.
It is easy to see that in this case, under time-reversal,
$$T[P^{0}] = -P^{0}, T[P^{1}] = P^{1}$$
while under parity
$$P[P^{0}] = P^{0}, P[P^{1}] = -P^{1} $$
This flows from the transformational properties of $x^{0}=t$ and x
under time-reversal and parity, respectively.
We also have that
$$[P^{0},P^{1}] = 0$$
Everything is as expected.
\chapter{Differential Equations on the Light-Cone  }
The Lagrangian in this case is
$$ \L = \partial_{+}\phi \partial_{-}\phi
  -{\lambda \over 4!} \phi^{4}$$
Following Chang, Root and Yan
\REF\CRY{S-J Chang, Robert G. Root and T-M Yan, Phys. Rev. \us{D7},
1133, (1973);
S-J Chang  and T-M Yan, Phys. Rev. \us{D7},1147, (1973). }
\refend
we define the canonical conjugate variable $\pi$ to $\phi$ thus
$$\pi={\partial \L \over \partial_{+}\phi} $$
if we define $x^{+}$ as the light-cone time.
This leads to the following modified hyperbolic
 differential equations of motion :
$$  \partial_{+}\partial_{-}\phi ={\lambda \over 3!}
\phi^{3}$$
Note that now this equation is linear in light-cone time. This means
that we get only one time
integration constant, hence only one boundary
condition, which is different from the equal-time quantization case .
It is not clear how to interpret the $x^{-}$ variable.

Another problem is that we don't recover the Hamilton's equations
of motion: we should have  two coupled equations
$$\partial_{+}\pi = \{ \H, \phi \} $$
and
$$\partial_{+}\phi = \{ \H, \pi \} $$
which upon integration over time
should give us two integration constants or two boundary
conditions.
Instead , we have only
$$\partial_{-}\pi = \{ \H, \phi \} $$
which gives us only one boundary condition. Probably this is different
physics.

\chapter{Naive Light-Cone Quantization and Parity }
Let us now define the energy-momentum tensor $T^{\mu\nu}$ in the
naive light-cone quantization case :
$$T^{\mu\nu} = \partial ^{\mu} \phi \partial ^{\nu} \phi - g^{\mu\nu}
\L $$
and
$$P^{\mu} = \int dx^{-} T^{\mu +}$$
In this case, this means that
$$P^{+} = \int dx^{-} (\partial^{+} \phi)^{2}$$
while
$$P^{-} = \int dx^{-} {\lambda \over 4!} \phi ^{4} $$
It is easy to see that in this case, under time-reversal,
$$T[P^{+}] \ne -P^{-}, T[P^{-}] \ne P^{+}$$
while under parity
$$P[P^{+}] \ne P^{-}, P[P^{-}] \ne -P^{+} $$
This flows from the transformational properties of $x^{+}$ and $x^{-}$
under time-reversal and parity, respectively. In this case, the
Hamiltonian $P^{-}$ is not an eigenstate of $\us{either}$ parity or
time-reversal. Our proposal to fix this problem is to consider
a new way to quantize on the light-cone.
\chapter{New Light-Cone Quantization }
The Lagrangian is
$$ \L = \partial_{+}\phi \partial_{-}\phi
  -{\lambda \over 4!} \phi^{4}$$
This leads to the following modified hyperbolic
second order differential equations of motion :
$$  \partial_{+}\partial_{-}\phi ={\lambda \over 3!}
\phi^{3}$$
Let us now treat the $^{+}$ and $^{-}$ variables on equal footing and
let us define now $\us{two}$ types of momenta:
$$p={\partial \L \over \partial(\partial_{-}\phi) } =\partial_{+}\phi$$
$$r={\partial \L \over \partial(\partial_{+}\phi) }=\partial_{-}\phi $$
Then there will be two types of Hamiltonians
$$ \H = 2 p \partial_{-}\phi  - \L = pr
+{\lambda \over 4!} \phi^{4} $$
and
$$ \K = 2 r \partial_{+}\phi - \L = rp
+{\lambda \over 4!} \phi^{4} $$
one, $\H$ for evolutions along $x^{+}$ and another, $\K$ for
evolutions along $x^{-}$.
We get now  coupled linear differential equations :
$$p=\partial_{+}\phi = \{ \H,r \}$$
and
$$r=\partial_{-}\phi = \{ \K,p \}$$
(these just reproduce the definitions above) as well as
$$\partial_{-}p=\partial_{-}\partial_{+}\phi=\{\K,\phi \} ={\lambda
\over 3!}\phi^{3} $$
$$\partial_{+}r=\partial_{+}\partial_{-}\phi=\{\H,\phi \} ={\lambda
\over 3!}\phi^{3} $$
which give the equation of motion.
The doubling of time variables would seem to imply the doubling
of boundary conditions, yet we want the same physics as in the
equal-time case. We have to show that some of the equations are
redundant.
Note that these last two equations
are identical, so should have  the same integration  constants,
or same boundary conditions.
This means that we'll have two boundary conditions, like in the
equal-time case.
\chapter{Parity Conserving Light-Cone Quantization and Parity }
Define now the energy-momentum tensor $T^{\mu\nu}$ thus
$$T^{\mu\nu} =   \partial ^{\mu} \phi \partial ^{\nu} \phi - g^{\mu\nu}
\L $$
We  have now the following   definition  for $P^{\mu}$
$$P^{\mu} = \int d\sigma_{\nu}T^{\mu \nu}$$
For our case where we treat $x^{+}$ and $x^{-}$ on equal footing, this
means that $d\sigma^{\nu}$ has $^{+} \us{and} ^{-}$ components :
$$P^{\mu} = \int dx^{-} T^{\mu -}+  \int dx^{+} T^{\mu +}$$
Just  like in McCartor's case, the generators contain
both $x^{+}$ and $x^{-}$ evolutions
\REF\gary{Gary McCartor, Z. Phys. \us{C41}, 271 (1988) .}
\refend  .
In this case, we get  the following  :
$$P^{+} = \int dx^{+}({\lambda \over 4!} \phi ^{4})
+\int dx^{-}(\partial^{+}\phi\partial^{-}\phi )$$
while
$$P^{-} = \int dx^{-}({\lambda \over 4!} \phi ^{4})
+\int dx^{+}(\partial^{-}\phi\partial^{+}\phi )$$
It is easy to see now that in this case, under time-reversal,
$$T[P^{+}] =  P^{-}, T[P^{-}] =   P^{+}$$
while under parity
$$P[P^{+}] = P^{-}, P[P^{-}] =  P^{+} $$
This flows from the transformational properties of $x^{+}$ and $x^{-}$
under time-reversal and parity, respectively. In this case, the
new definitions preserve both parity and time-reversal.
We also have the usual property that
$$[P^{+},P^{-}] = 0$$
like in the equal time quantization case.
\REF\phitreeplus{T. Heinzl, S. Krusche and E. Werner, 'Non-Perturbative
 Light Cone  Quantum  Field Theories Beyond the Tree Level,' TPR-92-16.}
\refend
\REF\phitreezero{T. Heinzl, S. Krusche and E. Werner, 'Zero Mode
 Corrections in Light-Cone  Quantum  Field Theory ,' TPR-92-17.}
\refend
\chapter{Constrained Light-Cone Quantization and Parity}
It might be interesting to check parity invariance in the context of
constrained light-cone quantization  \refmark\phitree,
\refmark\phitreezero, \refmark\phitreeplus,
\REF\clcq{Ovid C. Jacob, 'Constrained Light-Cone Quantization
Applied to 1+1-Dimensional Models:I. The Yukawa Model', SLAC-PUB,
June 1992 . }
\refend .
We can say even without doing any explicit calculations that both
parity and time-reversal will be broken since in that construction
we are really concerned with getting the correct $\H = P^{-}$, and
we do not worry about $P^{+}$, which is supposed to be a kinematic
generator anyway. The situation is just like in the naive light-cone
\chapter{Conclusions}
It is found that introducing two light-cone times, we recover
the appropiate number of boundary conditions to have
 the same type of theory in both equal-time and light-cone time.
This approach has the extra merit that it does not spoil parity
of time reversal
upon quantization, as does the naive light-cone quantization
approach.
\chapter{Acknowledgements}
I  would like to thank Dr. Dharam Ahluwalia for many useful discussions,
Prof. Stanley Brodsky for
discussions and to thank Prof. Richard Blankenbecler
for his continuing support; I  would also like to thank the organizers
of the Dallas Workshop on Light-Cone Quantization for putting together
such a simulating meeting which gave impetus to this work.
I also want to thank Prof. Michael Peskin for pointing out to me the
refernces on $N=2$ string quantization .
\refout
\end